# A Novel Radar Constant False Alarm Rate Detection Algorithm Based on VAMP Deep Unfolding

Haoyun Zhang[1], Chengyang Zhang[1], Xueqian Wang[1], Gang Li[1,*], Xiao-Ping Zhang[2]

[1]*Department of Electronic Engineering, Tsinghua University, Beijing, China*
[2]*Tsinghua Shenzhen International Graduate School, Tsinghua University, Shenzhen, China*
*Corresponding Author: gangli@mail.tsinghua.edu.cn

*Abstract* — **The combination of deep unfolding with vector approximate message passing (VAMP) algorithm, results in faster convergence and higher sparse recovery accuracy than traditional compressive sensing approaches. However, deep unfolding alters the parameters in traditional VAMP algorithm, resulting in the unattainable distribution parameter of the recovery error of non-sparse noisy estimation via traditional VAMP, which hinders the utilization of VAMP deep unfolding in constant false alarm rate (CFAR) detection in sub-Nyquist radar system. Based on VAMP deep unfolding, we provide a parameter convergence detector (PCD) to estimate the recovery error distribution parameter and implement CFAR detection. Compared to the state-of-the-art approaches, both the sparse solution and non-sparse noisy estimation are utilized to estimate the distribution parameter and implement CFAR detection in PCD, which leverages both the VAMP distribution property and the improved sparse recovery accuracy provided by deep unfolding. Simulation results indicate that PCD offers improved false alarm rate control performance and higher target detection rate.**
*Index Terms* — **VAMP, deep unfolding, radar, CFAR, PCD.**

## I. INTRODUCTION

Conventional high-resolution radars perform target detection based on fully sampled signals [1], which impose a significant burden on signal acquisition and transmission. Using compressive sensing (CS) algorithms, sub-Nyquist radar systems can recover the observation scene using fewer resources in time and frequency domains [2].

During the past few years, a variety of CS algorithms have been proposed, which can generally be divided into the following categories: 1) combinatorial algorithms [3–7]; 2) convex optimization algorithms [8–10]; 3) greedy algorithms [11–13]; 4) non-convex optimization algorithms [14–16]; and 5) iterative thresholding algorithms [17–19].

Despite their widespread use, combinatorial CS algorithms require a particular noise-free configuration in the measurements [20], while greedy CS algorithms require a greater number of measurements to ensure precise reconstruction. Furthermore, both the convex and non-convex CS algorithms are constrained by significant computational complexity, rendering them difficult to implement for large-scale CS problems [20, 21].

In recent years, the Approximate Message Passing (AMP) and Vector AMP (VAMP) algorithms, as iterative thresholding methods, have been proposed to solve CS problems [22–24]. According to the statistical physics principle, the Onsager correction is introduced to significantly improve the convergence speed of AMP. Following this advancement, VAMP has been proposed to generalize AMP to solve CS problems with a wider range of observation matrices, i.e., right-orthogonally invariant observation matrices. VAMP solves CS problems based on the Bayesian approach through belief propagation, maintaining a low computational complexity similar to that of AMP.

In recent studies, deep unfolding has been proposed to tune parameters in model-based algorithms through deep learning, thereby achieving improved performance [25]. Adapted into a deep unfolding framework, VAMP eliminates the need to design hyperparameters. In addition, VAMP deep unfolding can achieve accelerated convergence and improved recovery accuracy, in contrast to the conventional VAMP algorithm [26].

Despite the aforementioned advantages, VAMP deep unfolding modifies the parameter values in traditional VAMP algorithm, resulting in the distribution parameter of recovery error of non-sparse noisy estimation not directly obtainable through the method of traditional VAMP, which prevents its utilization in constant false alarm rate (CFAR) detection in sub-Nyquist radar system.

To implement CFAR detection under sub-Nyquist radar scenarios, a complex row-orthogonal debiased detector (CROD) is proposed by Na and Huang [27]. Nevertheless, this detector presents several limitations. Firstly, CROD is applicable only to CS algorithms based on the least absolute shrinkage and selection operator (LASSO) framework and is not well suited for VAMP deep unfolding, which relies on a Bayesian framework. Secondly, CROD necessitates accurate knowledge of the additive white Gaussian noise (AWGN) power in the environment, which is challenging to achieve.

To address the limitations mentioned above, a parameter convergence detector (PCD) is proposed based on VAMP deep unfolding in this paper. Compared to CROD, which only employs the sparse solution of the CS algorithm, PCD utilizes both the sparse solution and non-sparse noisy estimation to iteratively estimate the distribution parameter of the recovery error accurately. Additionally, the non-sparse noisy estimation of VAMP deep unfolding is also utilized to obtain the test statistic to implement CFAR detection. The simulation results demonstrate that PCD provides improved false alarm rate control performance and a higher target detection rate.

## II. SIGNAL MODEL

### A. VAMP Deep Unfolding

The complex received signal $y \in \mathbb{C}^{M \times 1}$ of the sub-Nyquist radar can be expressed as:

$$y = Ax_0 + n, \tag{1}$$

where $A \in \mathbb{C}^{M \times N}$ denotes the observation matrix, complex vector $x_0 \in \mathbb{C}^{N \times 1}$ denotes the observation scene to be recovered and $n \in \mathbb{C}^{M \times 1}$ denotes the complex additive white Gaussian noise (AWGN). The sparse recovery process using VAMP deep unfolding can be expressed as:

$$r_{R,I}, \hat{x}_{R,I} = V(y_{R,I}, A_{R,I}, K, \eta_{sst}), \tag{2}$$

where function $V(\cdot)$ represents the VAMP deep unfolding process. Here, $y_{R,I} = [\text{Re}(y) \ \text{Im}(y)]^T = [y_R \ y_I]^T \in \mathbb{R}^{2M \times 1}$ and $A_{R,I} = \begin{bmatrix} \text{Re}(A) & -\text{Im}(A) \\ \text{Im}(A) & \text{Re}(A) \end{bmatrix} = \begin{bmatrix} A_R & -A_I \\ A_I & A_R \end{bmatrix} \in \mathbb{R}^{2M \times 2N}$ are the decomposed forms of $y$ and $A$, respectively, since VAMP deep unfolding can only process data in real form. The notation $K$ represents the layer of VAMP deep unfolding and stage estimator $\eta_{sst}$ represents the scaled soft-thresholding function. The non-sparse noisy estimation $r_{R,I} = [r_R \ r_I]^T \in \mathbb{R}^{2N \times 1}$ denotes the pseudo-measurement and the sparse solution $\hat{x}_{R,I} = [\hat{x}_R \ \hat{x}_I]^T \in \mathbb{R}^{2N \times 1}$ denotes the denoised result. More details about VAMP deep unfolding are provided in Ref. [26].

### B. CFAR Detection

In CFAR detection task, we aim to detect targets in observation scene $x_0$ on an element-wise basis, which can be expressed as the hypothesis testing problem in the following:

$$\begin{cases} \mathcal{H}_{1,i}: x_{0,i} \neq 0 \\ \mathcal{H}_{0,i}: x_{0,i} = 0 \end{cases}, i = 1,2,\ldots,N, \tag{3}$$

where $x_{0,i} \neq 0$ and $x_{0,i} = 0$ represent the target is present and not present in the $i$th element of $\boldsymbol{x}_0$, respectively.

In VAMP deep unfolding, no additional nonlinear transformations are introduced compared to the conventional VAMP algorithm, thus maintaining the statistical distribution of the non-sparse noisy estimation $\boldsymbol{r}_{R,I}$, which follows a Gaussian distribution with the mean vector of $\boldsymbol{x}_{0,R,I} = [\text{Re}(\boldsymbol{x}_0) \ \text{Im}(\boldsymbol{x}_0)]^T \in \mathbb{R}^{2N \times 1}$ and a variance of $\sigma_{R,I}^2$, denoted as $r_{R,I,i} \sim \mathcal{N}(x_{0,R,I,i}, \sigma_{R,I}^2)$. Here, $r_{R,I,i}$ and $x_{0,R,I,i}$ represent the $i$th element of $\boldsymbol{r}_{R,I}$ and $\boldsymbol{x}_{0,R,I}$, respectively. Thus, the recovery error $\boldsymbol{w}_{R,I} = \boldsymbol{r}_{R,I} - \boldsymbol{x}_{0,R,I}$ of the non-sparse noisy estimation $\boldsymbol{r}$ also follows a Gaussian distribution with zero-mean and a variance of $\sigma_{R,I}^2$, denoted as $w_{R,I,i} \sim \mathcal{N}(0, \sigma_{R,I}^2)$. Here, $w_{R,I,i}$ represents the $i$th element of $\boldsymbol{w}_{R,I}$. The test statistic $\boldsymbol{r} \in \mathbb{R}^{N \times 1}$ to solve the hypothesis testing problem expressed as (3) in CFAR detection can be obtained as:

$$\boldsymbol{r} = \sqrt{\boldsymbol{r}_R^2 + \boldsymbol{r}_I^2}. \tag{4}$$

The notations $r_{H_1,i}$ and $r_{H_0,i}$ represent the $i$th element of $\boldsymbol{r}$ where the target is present and not present, respectively. Thus, $r_{H_0,i}$ follows a Rayleigh distribution [28], which can be expressed as:

$$p(r_{H_0,i}) = p(r_i|H_0) = \frac{r_i}{\sigma_{R,I}^2} \exp\left[-\frac{r_i^2}{2\sigma_{R,I}^2}\right], r_i > 0. \tag{5}$$

The false alarm rate, according to the Neyman-Pearson criterion, is given as:

$$P_{fa} = \int_{T_{P_{fa}}}^{+\infty} p(r_i|H_0) dr_i = \int_{T_{P_{fa}}}^{+\infty} \frac{r_i}{\sigma_{R,I}^2} \exp\left[-\frac{r_i^2}{2\sigma_{R,I}^2}\right] dr_i, \tag{6}$$

where $T_{P_{fa}}$ denotes the detection threshold when the false alarm rate is $P_{fa}$. According to equation (5), the relationship between $P_{fa}$ and $T_{P_{fa}}$ can be derived as:

$$T_{P_{fa}} = \sqrt{-2\sigma_{R,I}^2 \log(P_{fa})}. \tag{7}$$

According to the traditional VAMP, $\sigma_{R,I,\text{VAMP}}^2$ is obtained as the approximation of $\sigma_{R,I}^2$, which can be expressed as:

$$\sigma_{R,I,\text{VAMP}}^2 = \frac{\tilde{\sigma}_K^2 \tilde{v}_K}{1 - \tilde{v}_K}, \tag{8}$$

where $\tilde{\sigma}_K^2$ and $\tilde{v}_K$ are the intermediate variables in VAMP [24, 26]. However, the parameter constraints in traditional VAMP are altered by deep unfolding, which makes $\sigma_{R,I,\text{VAMP}}^2$ an inaccurate estimation of $\sigma_{R,I}^2$. To accurately estimate $\sigma_{R,I}^2$ and implement CFAR detection, we propose the PCD, which will be detailed in the following section.

## III. PROPOSED METHOD

After utilizing VAMP deep unfolding to complete sparse recovery, we can obtain the sparse solution $\hat{x}_{R,I}$, which acts as a reasonably precise approximation of the observation scene $x_{0,R,I}$. Therefore, in the first iteration in PCD, the approximation of $x_{0,R,I}$ can be obtained as:

$$\hat{x}_{R,I}^{(1)} = \hat{x}_{R,I}. \tag{9}$$

The element $w_{R,I,i}$ can be expressed as:

$$w_{R,I,i} = r_{R,I,i} - x_{0,R,I,i} \approx r_{R,I,i} - \hat{x}_{R,I,i}^{(1)}, \tag{10}$$

where $\hat{x}_{R,I,i}^{(1)}$ is the $i$th element of $\hat{x}_{R,I}^{(1)}$. For $i \in \{U \backslash \text{supp}(\hat{x}_{R,I}^{(1)})\}$, i.e., $\hat{x}_{R,I,i}^{(1)} = 0$, we can obtain:

$$w_{R,I,i} \approx r_{R,I,i} - \hat{x}_{R,I,i}^{(1)} = r_{R,I,i} - 0 = r_{R,I,i}, \tag{11}$$

where $\text{supp}(\hat{x}_{R,I}^{(1)}) = \{i: \hat{x}_{R,I,i}^{(1)} \neq 0\}$ represents the support set of $\hat{x}_{R,I}^{(1)}$ and $U = \{1,2,\ldots,N\}$ represents the universal set. Select the components corresponding to $i \in \{U \backslash \text{supp}(\hat{x}_{R,I})\}$ and arrange them into a vector $x_s \in \mathbb{R}^{L \times 1}$ where the number of the elements in the set $\{U \backslash \text{supp}(\hat{x}_{R,I})\}$, denoted as $L$, can be obtained as:

$$L = |U \backslash \text{supp}(\hat{x}_{R,I})|. \tag{12}$$

Then the power $\sigma_{R,I}^2$ estimated in the first iteration in PCD can be obtained as:

$$\sigma_{R,I}^{2\,(1)} = \frac{\sum_{i=1}^{L}\left(x_{s,i} - \frac{\sum_{i=1}^{L} x_{s,i}}{L}\right)^2}{L-1}, \tag{13}$$

where $x_{s,i}$ is the $i$th element of $x_s$. Consequently, in the first iteration in PCD, $w_{R,I,i}$ and $r_{R,I,i}$ can be approximately considered to follow Gaussian distributions, represented as $w_{R,I,i} \sim \mathcal{N}\left(0, \sigma_{R,I}^{2\,(1)}\right)$ and $r_{R,I,i} \sim \mathcal{N}\left(x_{0,R,I,i}, \sigma_{R,I}^{2\,(1)}\right)$, respectively. Therefore, $r_{H_0,i}$ can be approximately considered to follow a Rayleigh distribution, which can be represented as:

$$p(r_{H_0,i}) = p(r_i|H_0) = \frac{r_i}{\sigma_{R,I}^{2\,(1)}} \exp\left[-\frac{r_i^2}{2\sigma_{R,I}^{2\,(1)}}\right]. \tag{14}$$

Therefore, the detection threshold in the first iteration in PCD when the false alarm rate is set to $P_{fa_0}$ can be obtained as:

$$T_{P_{fa_0}}^{(1)} = \sqrt{-2\sigma_{R,I}^{2\,(1)} \log(P_{fa_0})}, \qquad (15)$$

where $P_{fa_0}$ is the appropriate false alarm rate selected to estimate the observation scene. The detection result in the first iteration in PCD can be obtained as:

$$\hat{x}_{P_{fa_0}}^{(1)} = r \odot I\left(r > T_{P_{fa_0}}^{(1)}\right), \qquad (16)$$

Where the notation $\odot$ represents Hadamard product and the element-wise index function $I(x)$ equals 1 when $x$ is true. Then the approximate estimation of the observation scene $x_{0,R,I}$ in the second iteration in PCD $\hat{x}_{R,I}^{(2)}$ can be obtained utilizing $\hat{x}_{P_{fa_0}}^{(1)}$, which can be expressed as:

$$\hat{x}_{R,I}^{(2)} = \left[r_R \odot I\left(\hat{x}_{P_{fa_0}}^{(1)} \neq 0\right); r_I \odot I\left(\hat{x}_{P_{fa_0}}^{(1)} \neq 0\right)\right]. \qquad (17)$$

Following the above steps, the estimate of $\sigma_{R,I}^2$ in the second iteration in PCD $\sigma_{R,I}^{2\,(2)}$ can be obtained. We repeat the process until $\sigma_{R,I}^{2\,(m)}$ converges, which can be expressed as:

$$\frac{\left|\sigma_{R,I}^{2\,(m)} - \sigma_{R,I}^{2\,(m-1)}\right|}{\sigma_{R,I}^{2\,(m-1)}} < c_{\text{tol}}, \qquad (18)$$

where the tolerance level $c_{\text{tol}}$ is a positive real value indicating convergence. We can obtain the power estimated by PCD $\sigma_{R,I,PCD}^2$ after the estimation $\sigma_{R,I}^{2\,(m)}$ converges, which can be expressed as:

$$\sigma_{R,I,PCD}^2 = \sigma_{R,I}^{2\,(m)}. \qquad (19)$$

Therefore, when the desired false alarm rate is $P_{fa}$, the threshold $T_{P_{fa}}$ can be obtained as:

$$T_{P_{fa}} = \sqrt{-2 * \sigma_{R,I,PCD}^2 * \log(P_{fa})}. \qquad (20)$$

Subsequently, the detection result $\hat{x}_{P_{fa}}$ can be obtained as:

$$\hat{x}_{P_{fa}} = r.* I\left(r > T_{P_{fa}}\right). \qquad (21)$$

The steps of PCD are detailed in Algorithm 1.

```
Algorithm 1: Parameter Convergence Detector
Input: P_{fa_0}, P_{fa}, c_{tol}, \hat{x}_{R,I}, r_{R,I}, r_R, r_I, r, m_{max}
Output: σ²_{R,I,PCD}, \hat{x}_{P_{fa}}
```
1. Initialize $\hat{x}_{R,I}^{(1)} = \hat{x}_{R,I}$
2. **for** $m \leftarrow 1$ **to** $m_{max}$ **do**
3.    $L = \left| U \backslash \text{supp}\left(\hat{x}_{R,I}^{(m)}\right) \right|$
4.    $x_s = r_{R,I}\left[U \backslash \text{supp}\left(\hat{x}_{R,I}^{(m)}\right)\right]$
5.    $\sigma_{R,I}^{2\,(m)} = \sum_{i=1}^{L}\left(x_{s,i} - \frac{\sum_{i=1}^{L}x_{s,i}}{L}\right)^2 / (L-1)$
6.    **if** $m \geq 2$ **then**
7.      **if** $\left|\sigma_{R,I}^{2\,(m)} - \sigma_{R,I}^{2\,(m-1)}\right|/\sigma_{R,I}^{2\,(m-1)} < c_{tol}$ **OR** $m = m_{max}$ **then**
8.        $\sigma_{R,I,PCD}^2 = \sigma_{R,I}^{2\,(m)}$
9.        $T_{P_{fa}} = \sqrt{-2 * \sigma_{R,I,PCD}^2 * \log(P_{fa})}$
10.       $\hat{x}_{P_{fa}} = r \odot I\left(r > T_{P_{fa}}\right)$
11.       **return** $\sigma_{R,I,PCD}^2, \hat{x}_{P_{fa}}$
12.      **end**
13.    **end**
14.    $T_{P_{fa_0}}^{(m)} = \sqrt{-2 * \sigma_{R,I}^{2\,(m)} * \log(P_{fa_0})}$
15.    $\hat{x}_{P_{fa_0}}^{(m)} = r \odot I\left(r > T_{P_{fa_0}}^{(m)}\right)$
16.    $\hat{x}_{R,I}^{(m+1)} = \left[r_R \odot I\left(\hat{x}_{P_{fa_0}}^{(m)} \neq 0\right); r_I \odot I\left(\hat{x}_{P_{fa_0}}^{(m)} \neq 0\right)\right]$
17. **end**

## IV. SIMULATION RESULTS

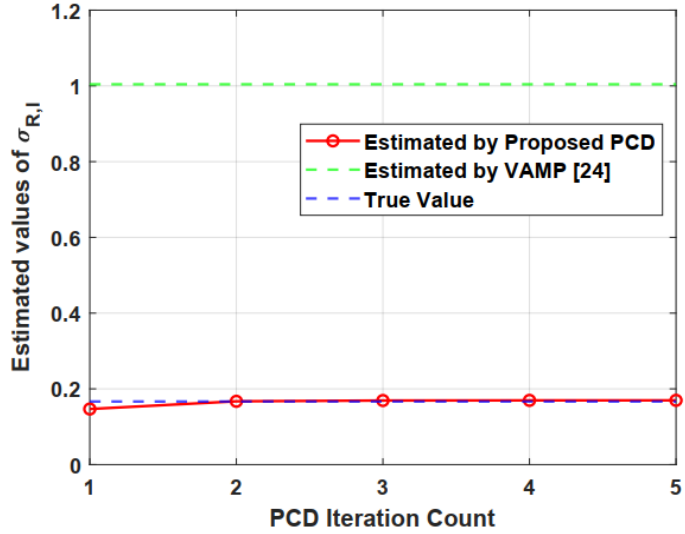

Fig. 1: Estimates of $\sigma_{R,I}$ obtained using different methods.

Fig.1. compares the standard deviation of the recovery error $\sigma_{R,I} = \sqrt{\sigma_{R,I}^2}$ estimated by PCD $\sigma_{R,I,PCD} = \sqrt{\sigma_{R,I,PCD}^2}$ and traditional VAMP $\sigma_{R,I,VAMP} = \sqrt{\sigma_{R,I,VAMP}^2}$. It is evident that $\sigma_{R,I,VAMP}$ significantly deviates from $\sigma_{R,I}$, as deep unfolding alters the parameter values in VAMP, rendering its traditional estimate invalid. In contrast,

$\sigma_{R,I}{}^{(m)} = \sqrt{\sigma_{R,I}^2{}^{(m)}}$ obtained from PCD rapidly converges to the true value $\sigma_{R,I}$. This result demonstrates the effectiveness of PCD in accurately estimating the recovery error power. That primary reason for this behavior is that, with an appropriate small $P_{fa_0}$, more and more $r_{R,i}$ and $r_{I,i}$ corresponding to $r_{H_0,i}$ are iteratively incorporated into $x_s$. Consequently, the gap between $\sigma_{R,I}^2{}^{(m)}$ and $\sigma_{R,I}^2$ progressively decreases. The observation matrix is partial Fourier observation matrix with dimension $M = 600, N = 1000$. We set $P_{fa_0} = 10^{-5}, c_{\text{tol}} = 10^{-5}$ in PCD and conducted 10,000 Monte Carlo experiments to obtain the ECDFs as illustrated in Fig. 1.

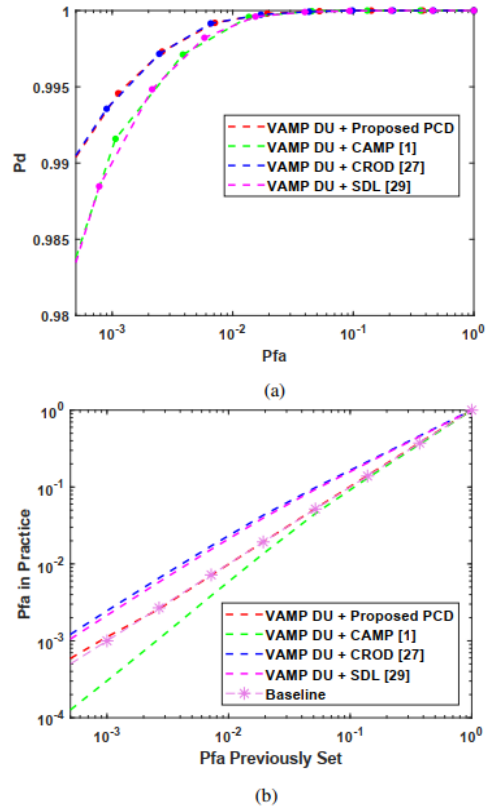

Fig. 2: (a) Receiver Operating Characteristic (ROC) curves and (b) False alarm rate control for VAMP deep unfolding with detectors PCD, CAMP [1], CROD [27], SDL [30].

Fig.2. compares the performance of various different detectors combined with VAMP deep unfolding. It can be observed that the detection rate $P_d$ of VAMP deep unfolding combined with PCD is comparable to that achieved with CROD and is significantly superior to that with CAMP[1] and SDL. In particular, while CROD acquires more prior knowledge than PCD to implement detection, PCD demonstrates the best false alarm

control performance of all detectors. The reason for this behavior is that PCD leverages both the sparse solution and non-sparse noisy estimation in VAMP deep unfolding, which fully exploits the distribution property of VAMP and the improved sparse recovery accuracy provided by deep unfolding.

Fig.3. compares the performances of different CS algorithms combined with their optimal detectors. It demonstrates that VAMP deep unfolding with PCD outperforms competing methods in target detection and better false alarm control. It can be concluded that PCD effectively translates the enhanced accuracy offered by deep unfolding into superior sub-Nyquist radar CFAR detection performance. The observation matrix is partial Fourier observation matrix with dimension $M = 200, N = 1000$. We set $P_{fa_0} = 10^{-3}, c_{\text{tol}} = 10^{-20}$ in PCD and conducted 1,000 Monte Carlo experiments to obtain the results in Fig. 2 and Fig. 3.

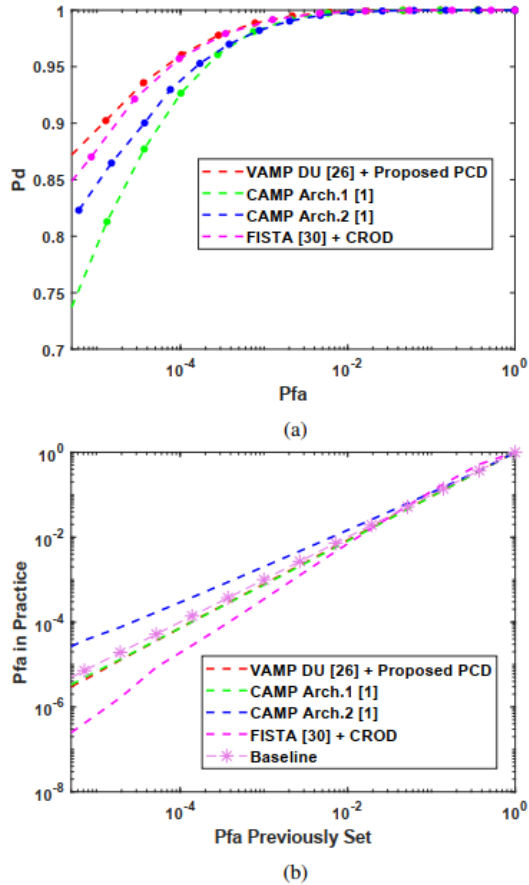

Fig. 3: (a) ROC curves and (b) False alarm rate control for CAMP [1] and VAMP deep unfolding [26], FISTA [30] with their optimal detectors.

V. CONCLUSION

In this paper, we propose PCD to implement sub-Nyquist CFAR detection based on deep unfolding. PCD leverages both the VAMP distribution property and the enhanced accuracy provided by deep unfolding, which achieves improved target detection performance and false alarm control performance than the state-of-the-art approaches. Additionally, compared to CROD, PCD necessitates no prior information about the power of AWGN, making it more suitable for practical application. In conclusion, PCD bridges the gap between the enhanced accuracy offered by deep unfolding and the superior radar CFAR detection performance.

[1]CAMP [1] is an integrated approach for CS and detection, and can also be used independently as a detector, as mentioned in Ref. [27].

ACKNOWLEDGMENT
This work was supported by the National Natural Science Foundation of China under Grant 62388102.